# I-V characteristics of in-plane and out-of-plane strained edge-hydrogenated armchair graphene nanoribbons

S. J. Cartamil-Bueno,[1,a] and S. Rodríguez-Bolívar[2,a]

[1]*Kavli Institute of Nanoscience, Delft University of Technology, Lorentzweg 1, 2628 CJ Delft, The Netherlands*

[2]*Department of Electronic and Computer Technology, Faculty of Sciences, University of Granada, Campus Fuentenueva s/n, E-18071 Granada (Spain)*

The effects of tensile strain on the current-voltage (I-V) characteristics of hydrogenated-edge armchair graphene nanoribbons (HAGNRs) are investigated by using DFT theory. The strain is introduced in two different ways related to the two types of systems studied in this work: in-plane strained systems (A), and out-of-plane strained systems due to bending (B). These two kinds of strain lead to make a distinction among three cases: in-plane strained systems with strained electrodes (A1) and with unstrained electrodes (A2), and out-of-plane homogeneously strained systems with unstrained, fixed electrodes (B). The systematic simulations to calculate the electronic transmission between two electrodes were focused on systems of 8 and 11 dimers in width. The results show that the differences between cases A2 and B are negligible, even though the strain mechanisms are different: in the plane case, the strain is uniaxial along its length, while in the bent case the strain is caused by the arc deformation. Based on the study, a new type of NEMS-solid state switching device is proposed.

## I. INTRODUCTION

The miniaturization of devices is one of the main driving forces that has actively led research into the nanoscale regime. In nanoelectronics, the physical limits of the classical semiconductor technologies are being reached[1], and Moore's law will no longer work unless other materials than silicon and new computing principles are found. Furthermore, there exist several industrial and social constrains such as planar compatibility and environmental awareness that set the suitability of the possible substitutes for the technological change.

Graphene, a single atomic layer of carbon atoms arranged in a hexagonal (honeycomb) lattice isolated in 2004[2], could be the ultimate material compatible with the actual schemes as it is planar plus robust, stable, and combines exceptional electronic and mechanical properties[3–5]. However, it lacks of an energy band gap[6], only making it useful for analog applications or new types of digital devices[7]. Nevertheless, its discovery boosted the investigation on many other van der Waals materials such as molybdenum disulfide and boron nitride that could substitute silicon and its oxide[8–12], hence settling the field of two-dimensional materials. Derivative structures of those 2D materials such as graphene nanoribbons (GNRs) – 1D materials similar to carbon nanotubes but planar–[13–16] have attracted recently the interest of the Research Community as they involve new fundamental effects that could be used for novel functionalities, and will have the last word on the size minimization. Recently, the first experiments on under-10-nm-in-width GNRs[17–23] brought the first opportunities to investigate their electronic transport properties, although many hitches such as disorders (charge impurities and edge roughness) and quantum phenomena (Anderson localization and Coulomb blockade effect) could have prevented of obtaining the results predicted by theory[24]. However, this has not discouraged researchers from fabricating the first GNRFETs[25], and controlling the GNR width with atomic precision[26,27].

Nonetheless, the crystallography of graphene is known since much time before its isolation as it was used as a theoretical model[28]. Moreover, the first studies on GNRs date from late 1980s, and report the electronic states for the two different edge terminations (semiconducting and metallic for armchair and zigzag, respectively) via tight binding calculations[29–32], density functional theory[33], and the extended Hubbard model[34]. More recently, the dependence of the energy gap with the ribbon width in armchair-edge type nanoribbons (AGNRs) has been found to lie on the change of the bonding distance of the edge atoms[35].

As the technology is making possible the fabrication of this type of systems by means of either overcoming the top-down fabrication limits or developing efficient and reproducible bottom-up methods, more theoretical studies can be done on nanoribbons. There is an increasing interest in examining their mechanical[36–47] and electronic properties[48–58] under different constrains such as tensile stress to predict the behavior of future possible devices. Particularly, strain studies on graphene nanoribbons[59–82] are gaining much attention as the control of their mechanical deformation could allow the creation of novel devices for energy harvesting[83–86]. There exist many reports on the electronic structure modification via strain engineering of simulated systems of different edge type, edge decoration, shape, and width[46–50,87–96]. Among those researchers, some outstand by including analyses of the current-voltage characteristics[43,44,92,97,98]. However, to the best of our best knowledge, there is no literature about the changes in the electronic properties due to the dynamics of a suspended graphene nanoribbon. Indeed, we believe that a new type of switching device can be proposed: a ultimate-piezoresistive nanoelectromechanical system (NEMS) made out of a material whose electronic structure changes drastically and cyclically with the strain produced by its own motion; that is, a material that could change from metal/semimetal to semiconductor/dielectric and viceversa, within a single oscillation. Therefore, its switching mechanism would be intrinsically of solid-state nature instead of mechanic (physical contact) or electric (charge accumulation).

With this aim, in this work we use the DFT-based TranSIESTA program[99] to investigate the effects of tensile strain on the current-voltage (I-V) characteristics of two hydrogenated-edge armchair graphene nanoribbons (of 8 and 11 dimers in width) when subjected to an out-of-plane bending. We compare the results to the same study when different in-plane uniaxial strains are applied to the system.

The present article is divided in three sections: first, the theoretical model used is explained together with the description of the simulations and calculations with SIESTA and its TranSIESTA module; then, the results are analyzed and discussed; and finally, last section enumerates the conclusions.

## II. THEORETICAL MODEL AND SIMULATION PROCESS

The calculation of the electron transport in a finite system with the SIESTA package involves two steps: the SIESTA simulation of two electrodes through which a voltage will be applied; and the TranSIESTA simulation of the final system (also known as 'scattering region') composed of a central region and a few layers of the two electrodes (Fig. 1a-b). In this work, the simulations emulate graphene nanoribbons of defined width and length without other electrodes than its own extremities. Each electrode is an identical SIESTA cell consisting of two lines of carbon atoms plus hydrogen terminations, and the central region that they enclose is made out of five of those cells. Thus, the final system comprises seven electrode cells.

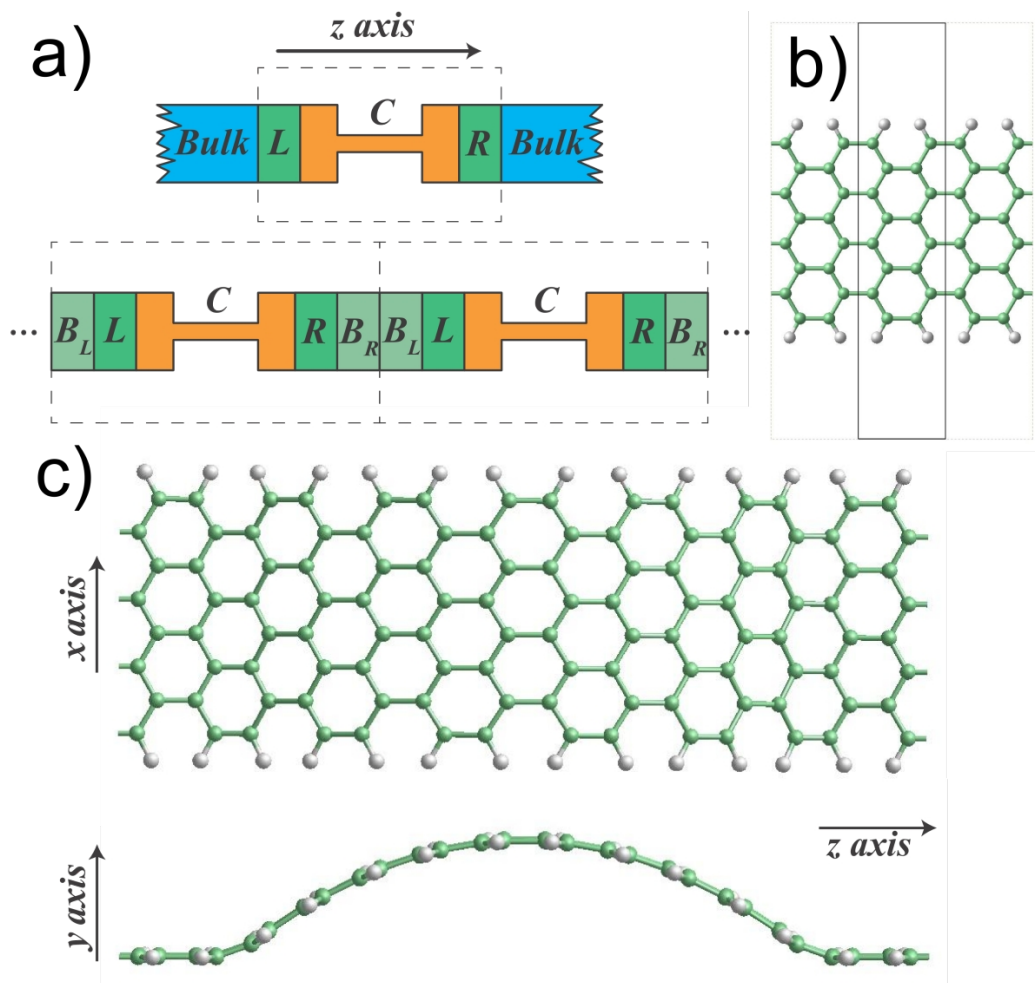

FIG. 1: Model of the system. (a) Conceptual illustrations: the definition of the cell that contains the device is represented in the top image. The scattering region consist of part of the –still infinite-long– left (L) and right (R) electrodes and the contact region (C). The system is made out of the repetition of the device cell suitable for a TranSIESTA simulation (bottom image): the electrodes become finite, but optional buffer layers (BL, BR) can be inserted between two consecutive cells. (b) The cell of an unstrained hydrogenated-edge armchair graphene nanoribbon (HAGNR) composed of 8 carbon atoms in width. The cell is repeated filling the whole 3D space, but only cells in the z direction connect among each other. (c) Two projections of a bent HAGNR of 8 atoms in width. Atomic visualizations obtained with the GDIS software.

In the first stage, SIESTA simulates an auxiliary bulk electrode which is made out of an user-created electrode cell. For a N-dimer-wide HAGNR, that electrode cell is composed of 2N carbon atoms and 4 hydrogen atoms. They are arranged in such a way that the adjacent reproduction of the cell over the *xz* plane produces infinite series of identical HAGNRs of infinite length in the *z* axis with sufficient space among them in the *x* axis to avoid interaction. Also, SIESTA requires to set the height of the cell (*y* axis) large enough to prevent interaction among layers. In this work, both the inter-HANGR and the inter-layer distances were set to 10 Å.

One of those HAGNRs is the so-called bulk electrode, and its electronic properties will be applied to the finite electrodes of the final system. As observed in Fig. 1c for the case of a 8-HAGNR, the atoms in the electrode cell initially have the ideal honeycomb structure of graphene with bond lengths set by a lattice constant of 2.485 Å (determined by minimizing the total energy of a graphene system).

In this step, we used single-zeta basis plus polarization(SZP) basis size and mixed functionals (GGA-PBE for C atoms, LDA-CA for H atoms) together with the following well-converged parameters: 300 Ry for mesh cutoff energy, and 45 k-points in the $k_z$ direction. Furthermore, the coordinate optimization method by conjugate gradients (CG) was used for atomic relaxation (maximum tolerance of 0.01 eV/Å in the force balance), which was combined with a variable cell scheme to perform a 'lattice relaxation' (maximum stress tolerance of 0.0001 eV/Å$^3$ for a target pressure of 0 GPa). Despite aiming for zero stress, the relaxation always returns a finite stress that we report for each case in the results section. This double relaxation is of great importance as it will be discussed later: although bulk graphene and graphite maintain the structure of ideal hexagons, nanoribbon edge atoms tend to collapse into the body of the system causing an anisotropic deformation of the hexagons[35].

In the second stage, the relaxed atomic positions from the electrode cell are used to make the final system; and together with the calculated electrode properties, TranSIESTA calculates the electronic transmission spectra for different voltages at a determined temperature (300 K in this paper).

The width and zero-strain state of the final HAGNR are defined by the width and the minimum energy atomic configuration of the electrode from the previous stage, respectively. The final length is set by the number of electrode cells that the final system contains, and the applied strain. A length of seven electrode cells is chosen to conform the final system for commodity reasons since the electronic transmission should not be affected as long as the central region is neither too short to cause electrode-to-electrode interference or confinement, nor longer than the electron mean free path. The strain is introduced in two different ways related to the two types of systems studied in this paper: in-plane strained systems (A), and out-of-plane strained systems due to bending (B). Figure 1c shows two projections from the same bent system.

For the final system simulations, we used SZP basis size and mixed functionals (GGA-PBE for C atoms, LDA-CA for H atoms) together with the following well-converged parameters: 300 Ry for mesh cutoff energy, and 3 k-points in the $k_z$ direction. In this step, no coordinate optimization method (relaxation) was performed. No buffer atoms were used.

Lastly, the results from both calculations are used by the TBTrans program to extract the transmission spectrum and the electronic current.

## A) Plane systems

These HAGNRs maintain all atoms in the same *xz* plane, but their cell dimensions are changed via the lattice vector parameters. In SIESTA, the atomic coordinates can be set relatively to the lattice vectors ("AtomicCoordinatesFormat fractional"). Thus, as the strain is scalable, by multiplying the lattice vector in the *z* direction, all atom-atom distances in that direction are modified in the same proportion, hence emulating a homogeneous uniaxial strain. This 'engineering normal strain' or 'nominal strain' ε is defined as

$$ \qquad (1)$$

where *L* is the original length (7 times the length of the relaxed electrode cell in this work) and *l* is the final length.

Furthermore, two systems of this kind are studied in this report: HAGNRs whose electrodes have the same strain than in the central region, and HAGNRs whose electrodes are unstrained while the rest of the structure is stretched.

## B) Bent systems

These HAGNRs bend upwards but maintaining the electrode atoms fixed in the same positions. Thus, the atoms from the central region are no longer delimited to the *xz* plane, and they move to positions of different heights (in the *y* direction) hence forming an arc. The strain arises from imposing a larger atom path for the same amount of atoms; however, two types of strains can be defined: homogeneous or inhomogeneous. The former considers the atoms being arranged with equidistant positions, while the latter set the atoms from the plane to the arc positions by moving them only in the *y* direction (which results in lower strain right in the middle of the structure, and very large strain between the electrode and the neighbor atoms at the beginning of the central region). In principle, we believe that nature would choose the homogeneous strain as it would lead to lower energy and, that is why, we only simulate it; however, it is possible that in certain experiments, the inhomogeneous strain happens (*e.g.*, due to electrostatic coupling by a strong and normal electric field, or due to clamping).

Beside assuming homogeneous strain, we make the hypothesis that the central region forms an actual parabola

$$ \tag{2} $$

where $y_i$ and $z_i$ are the *y* and *z* coordinates of the $i^{th}$ atom, h is the height or deflection, and *L* is the length of the central region when flat (zero curvature). The strain is defined again as

$$ \tag{3} $$

where *l* is the length of the arc. To calculate *l*, it is necessary to solve analytically the following line integral:

$$ . \tag{4} $$

### III. RESULTS

The first systematic simulations were performed to calculate the electronic band structure of HAGNRs of different widths and strains. For this purpose, only simulations of the electrodes were enough. The calculations were performed for two basis sizes: SZP and DZP (double-zeta basis plus polarization). Figures 2a-f show the calculated band structure for zero-strained HAGNRS of widths from 3 to 8 dimers (SZP basis size), and Fig. 2g compares the dependence of the band gap energy as function of the width for both basis sizes. As observed, there exist three distinct series of nanoribbons regarding the number of dimers in width in agreement with theory and other reports: 3m, 3m+1, and 3m+2 [48–58]. Their band gap energies decrease with the width and decay to zero as one could expect by recalling the fact that bulk graphene is a semimetal. Each series has a different initial band gap energy and they seem to have a different rate of convergence as well. Series 3m (3, 6, 9, 12,...) and 3m+1 (4, 7, 10, 13,...) are very similar: systems of the latter have larger band gap energies, although the difference becomes smaller gradually. Both are interesting for semiconductor applications as the band gaps of the thinner cases are of the order of the silicon one (1.1 eV). The 3m+2 series (5, 8, 11, 14,...) comprehend the ribbons of lowest band gap energies, and soon the systems becomes metallic. The different series and obtained bandgap values are very similar to those calculated with the same method and using the local density approximation (LDA) by Son et al.[35], in contrast to the different results by means of tight-binding from the same reference. On the technical side, it can be observed that calculations with DZP give appreciably larger band gaps for series 3m and smaller ones for 3m+2 systems, while there are no big differences for the case of 3m+1. However, this behavior is not hold when the strain is applied. In any case, no statements can be done regarding the advantages of using DZP. That is why, a SZP basis size was chosen for the transmission simulations as it provides enough quality for qualitative studies.

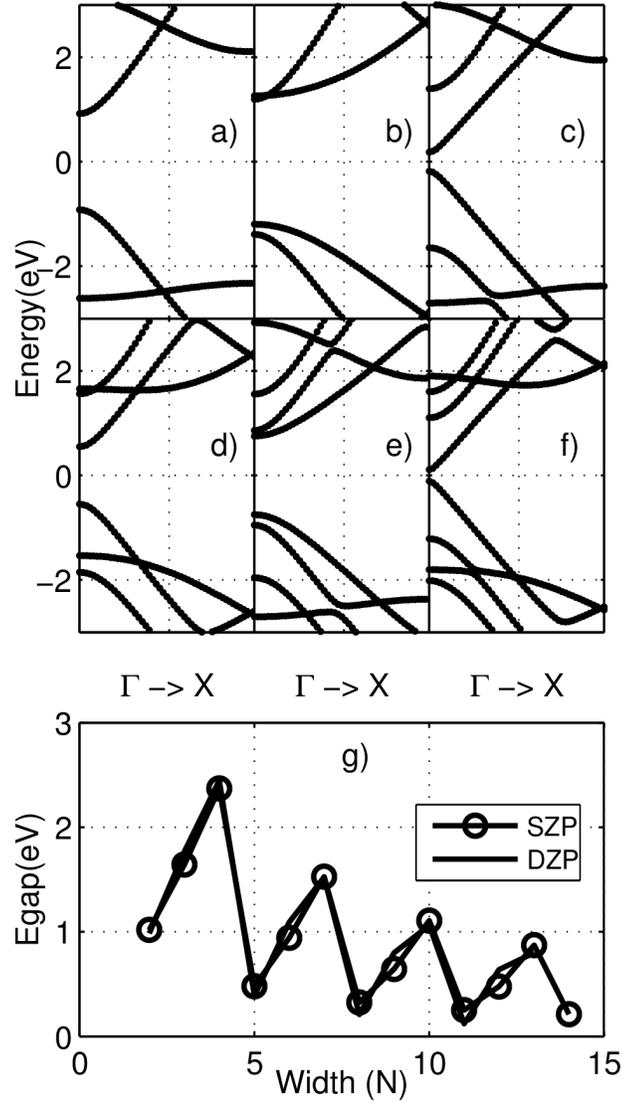

FIG. 2: Width study on unstrained HAGNRs. Energy band structure (SZP) of (a) 3-HAGNR, (b) 4-HAGNR, (c) 5-HAGNR, (d) 6-HAGNR, (e) 7-HAGNR, and (f) 8-HAGNR; and (g) the dependence of the band gap energy Eg as function of the number of dimers in width N (for SZP and DZP basis sizes). As observed, there exist three distinct series of nanoribbons regarding the number of dimers in width in agreement with theory and other reports: $3m$, $3m+1$, and $3m+2$. Their band gap energies decrease with the width and decay to zero as one could expect by recalling the fact that bulk graphene is a semimetal. Furthermore, calculations with DZP give appreciably larger band gaps for series $3m$ and smaller ones for $3m+2$ systems, while there are no big differences for the case of $3m+1$.

Figure 3 gathers the energy gap-strain graphs in different collections of series for the SZP case. It shows that there exist extrema points or 'kinks' for the band gap energy at particular strains. This peculiar dependence is also found in the literature[46–50,87–96]. It is noticed that as the systems become wider, new extrema points appear; that is to say, the abrupt changes happen at smaller strains together with the corresponding decrease in maximum band gap energy (maxima points become lower).

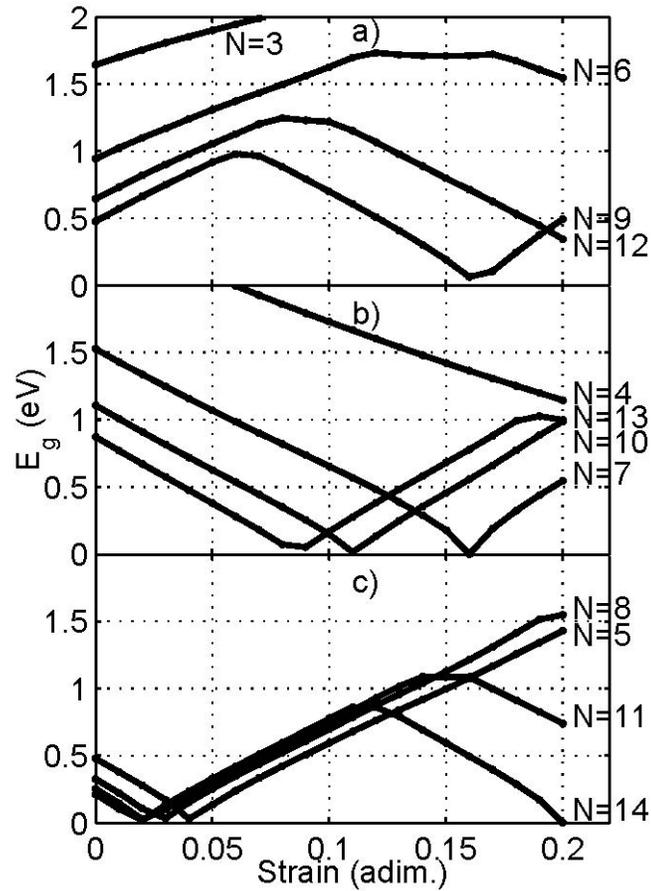

FIG. 3: Strain study on HAGNRs. Band gap energy $E_g$ as function of the strain $\varepsilon$ for HAGNRs of the (a) 3m series (N = 3, 6, 9, 12), (b) 3m+1 series (N = 4, 7, 10, 13), and (c) 3m+2 series (N = 5, 8, 14). Peculiar extrema points or 'kinks' exist at particular strains. As the systems become wider, those abrupt changes happen at smaller strains together with the corresponding decrease in maximum band gap energy. Band gap energy values were calculated with a SZP basis size.

Afterwards, the systematic simulations to calculate the electronic transmission between two electrodes were focused on systems of 8 and 11 dimers in width. As said in the previous section, the two kinds of strain lead to make a distinction among the resulting different systems: in-plane strained systems with strained electrodes (A1) and with unstrained electrodes (A2), and out-of-plane homogeneously strained systems with unstrained, fixed electrodes (B).

A1) Plane HAGNRs with strained electrodes

The I-V characteristics for 8-HAGNRs with strains applied to both the electrodes and the central region ranging from 0 to 20% are shown in Fig. 4. If our results are crosschecked with the work of Topsakal et al.[92], it has to be noticed that our calculated currents are the double than the reported values. That factor of 2 was found out to be an already-solved issue of the software[100].

Besides those doubled values, the gap energy curves for the different strains seem displaced by a strain factor when compared to the same reference. It may be explained by assuming that there is a zero-strain reference mismatch due to a different relaxation method: if the systems were not relaxed or were relaxed with different stress target or tolerance, the zero-strain system would actually have a random offset strain. In that case, an unstrained system in the present study is equivalent to another one with a certain intrinsic strain, and the data matches by applying a correcting factor in the strain. Figure 4a shows the dependence of the current with the strain for bias voltages from 0 V to 1.8 V at 300 K. When looking in detail for low strains and voltages, it is observed that the current from relaxed to a 3% strain increases drastically, about 8 orders of magnitude (Fig. 4b). The I-V curves in Fig. 4c indicate that the system has an ohmic behavior (13.4 kΩ) at the strain that corresponds to the lowest band gap energy (3%), while strains further than that produce a nonlinear behavior although still symmetric together with a threshold voltage that seems to be related to the gap energy.

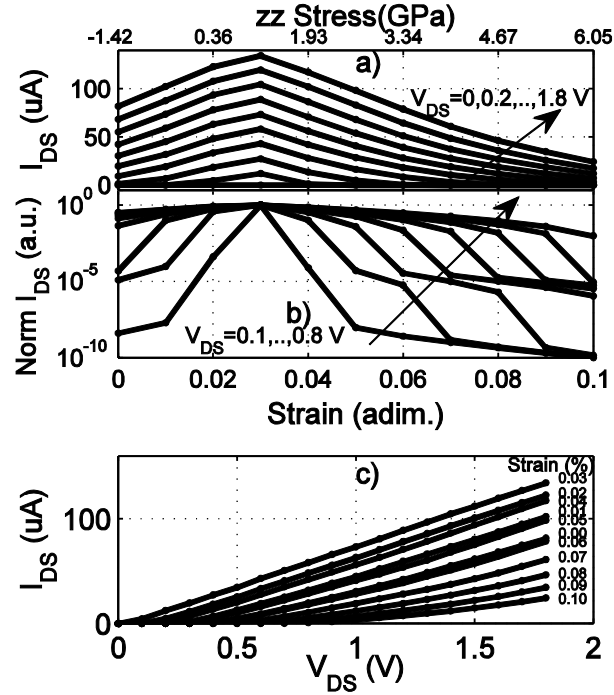

FIG 4: Study on plane HAGNRs of 8 dimers in width with strained electrodes. (a) Dependence of the current with the strain for bias voltages from 0 V to 1.8 V (T = 300 K). (b) Detail of the dependence with normalized currents for low strains and bias voltages. For a drain-source voltage of 0.1 V, the current increases more than 8 orders of magnitude when the system is stretched 3%. (c) I-V characteristic curves for different strains. In (a) and (b) the top axis is the *zz* component of the stress tensor. The minimum values for both the Total Static Pressure and the zz Stress happen for a strain of 2%, being -0.1358 GPa and 0.3642 GPa, respectively.

The same study for 11-HAGNRs is shown in Fig. 5. This study is not found in the literature, and no comparisons can be done. It can be recalled that 11-HAGNRs belong to the same series than 8-HAGNRs, and therefore both systems have a similar dependence with the strain. Indeed, as the kink strain happens at a smaller strain for 11-HAGNRs, the current peak is also displaced to a lower strain (2%). Furthermore, the maximum value of the current is about the same because the band gap energy is of the same order. In the literature, a note on the kink strain on 11-HAGNR simulations indicates that that effect is related to the Peierls effect[80]. Figure 5a-b show the current-strain dependence and the same dependence with normalized currents, respectively. The I-V characteristic curves for different strains are shown in Fig.5c.

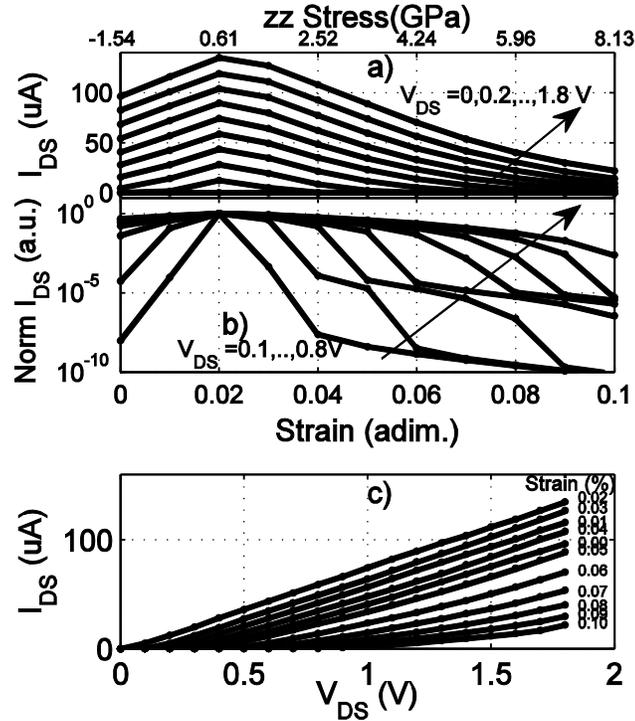

FIG 5: Study on plane HAGNRs of 11 dimers in width with strained electrodes. (a) Current-strain dependence for bias voltages from 0 V to 1.8 V (T = 300 K). The increase in the current at high strains is coherent with the calculated "peak kink" of the band gap energy observed in Fig. 3c. (b) Same dependence with normalized currents for low strains and bias voltages. The current peak occurs at 2% of strain for 0.1 V, and the corresponding Ipeak/Iunstrained ratio is 8 orders of magnitude. (c) I-V characteristic curves for different strains. In (a) and (b) the top axis is the *zz* component of the stress tensor. The minimum values for both the Total Static Pressure and the zz Stress happen for a strain of 1%, being 0.1326GPa and -0.4295GPa, respectively.

A2) Plane HAGNRs with unstrained electrodes

The I-ε and I-V characteristics for HAGNRs with the strain applied only to the central region while keeping the electrodes unstrained are shown in the left panels of Fig. 6-7 together with the curves of bent HAGNRs (right panels) for the sake of comparison. The comparative results will be discussed in the next section. The variation of the current with low strains for different bias voltages in a 8-HAGNR is shown in Fig. 6a. The maximum current is lower than in the system from the previous chapter, and it can only be due to the electrodes. Figure 6b shows its I-V characteristic curves, which are different to the plane systems with strained electrodes: there is no strain in which the system is purely ohmic, although all curves have the same threshold voltage (no longer proportional to the gap energy) and the ohmic regime lead to the same resistance (14.0 kΩ). In a same way, Fig. 7 exposes the corresponding curves for the 11-dimer-in-width case. For this case, the differences between the systems are again negligible.

B) Bent systems with homogeneous strain and unstrained electrodes

As initiated in the previous section, Fig. 6-7 (right panels) show the characteristic curves for 8- and 11-HAGNRs with different bending curvatures that causes an intrinsic homogeneous strain in the central region (the electrodes are fixed and remain unstrained). The results are of great interest as they show that the system behaviors are quite similar to the plane versions, even though the strain mechanisms are different: in the plane case, the strain is uniaxial along its length, while in the bent case the strain is caused by the arc deformation. It is interesting to know that a strain of 3% leads to a deflection of 2.31 Å for a 8-HAGNR of 21.52 Å in length (length/deflection ratio of 9.32). Similarly, in the 11-dimer-in-width case, the bending strain of 2% corresponds to a deflection of 1.88 Å for the same length (length/deflection ratio of 11.45).

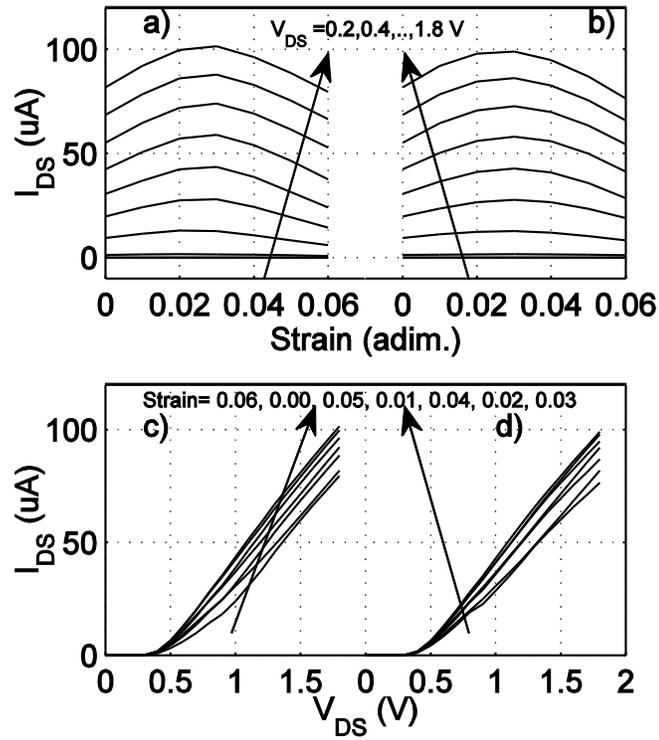

FIG 6: Comparison between a plane 8-HAGNR system with unstrained electrodes (left panels) and a bent 8-HAGNR system with homogeneous strain and unstrained electrodes (right panels). (a) Variation of the current with low strains for different bias voltages (T = 300 K). (b) I-V characteristic curves. The differences between the systems are negligible, even though the strain mechanisms are different: in the plane case, the strain is uniaxial along its length, while in the bent case the strain is caused by the arc deformation. In the latter, a strain of 3% corresponds to a deflection of 2.31 Å for a 21.52 Å in length (L/h ratio of 9.32).

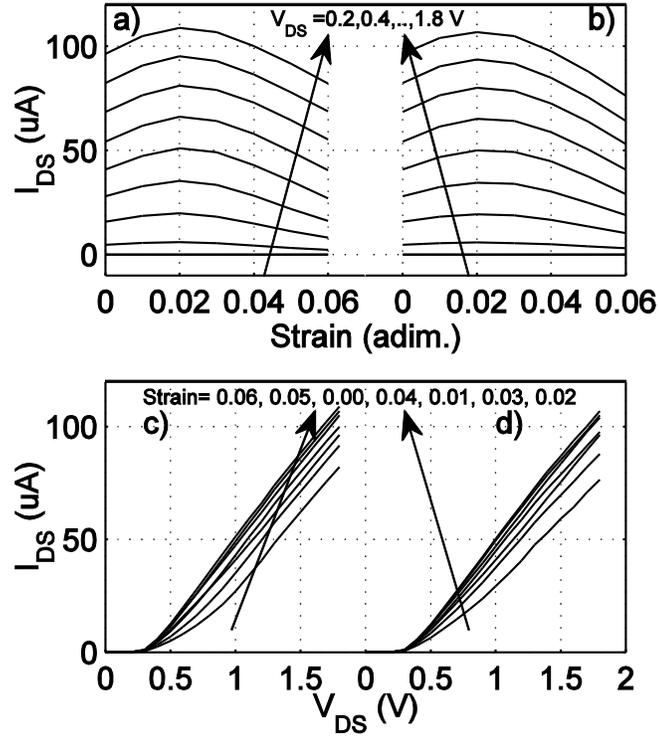

FIG 7: Comparison between a plane 11-HAGNR system with unstrained electrodes (left panels) and a bent 11-HAGNR system with homogeneous strain and unstrained electrodes (right panels). (a) Current-strain dependence for different bias voltages (T = 300 K). (b) I-V characteristic curves. The differences between the systems are again negligible. The bending strain of 2% corresponds to a deflection of 1.88 Å for a 21.52 Å in length (L/h ratio of 11.45).

**IV. CONCLUSIONS**

The effect of tensile strain on the current-voltage characteristics of hydrogenated-edge armchair graphene nanoribbons of 8 and 11 dimers in width has been studied by introducing the strain in two ways in this work: in-plane strained systems (A), and out-of-plane strained systems due to bending (B). These two kind of strains led to make a distinction among three cases: in-plane strained systems with strained electrodes (A1) and with unstrained electrodes (A2), and out-of-plane homogeneously strained systems with unstrained, fixed electrodes (B). A system from case A1 could correspond in the real world to a HAGNR fixed to a substrate which deforms such as the channel of a GNRFET on a flexible substrate. In contrast, a case A2 could be related to a suspended HAGNR of a particular pretension or clamped over a tunable trench. Despite being identical systems though treating the electrodes differently, it has been demonstrated that the I-V characteristic curves from cases A1 and A2 are appreciably different.

On the other hand, case B and case A2 give systems of very similar behavior, even though the strain mechanisms are different: in the plane case, the strain is uniaxial along its length, while in the bent case the strain is caused by the arc deformation. Results from case B systems could be interpreted from a different point of view. Instead of systems of different deflections and strains, each system of different curvature may correspond to a different state of the same system when oscillating as a double-clamped HAGNR resonator. The study of such a system is of big interest as it may become a new type of switching device which has the advantages of NEMS and solid-state transistors. Paying attention to Fig.3, one could dear to think that if a system with a determined width that gives a good compromise between kink strain (as smaller as possible) and band gap energy (as large as possible) could be engineered, this behavior could be exploited to fabricate NEMS switches with extremely high $I_{on}/I_{off}$ ratios that are not limited by the Boltzmann temperature. Besides being high performance oscillators and intrinsic frequency multipliers, for example, such a switching resonator could be turned into a revolving-bandgap transistor by controlling its resonant frequency with a third terminal, and pave the way for energy/optical harvesting.

It is difficult to compare the results from this work to others from literature because the actual straining of the system is not reported with clarity. We suspect that most of the works with calculations on the current-voltage characteristics of HANGRs[43,44,92,97] are based on systems A1 (strained electrodes). Only references 43, 92 and 97 study a system of 8 atoms in width, while no literature was found for HAGNR of 11 atoms in width (although it belongs to the same series of 8 atoms and 5 atoms in width, also studied in reference 44). Our results from Fig.4 are qualitatively similar to the three references before. Quantitatively, the values of the current, bias voltage and strain match better the results from reference 97. No reference was found to plot the current versus the strain, a characteristic curve that we believe of great interest as it clearly visualizes the suitability of the system for strained electronic devices.

Finally, it should be remarked that relaxing a system is a crucial process that is underestimated very often and typically not reported. Depending on how the relaxation is performed, a different system than the studied one could be actually being simulated. Discouragingly, it is suspected that many of the reported calculations come from simulations where either just the electrode cell, just the scattering region or both of them were either not relaxed at all or severely relaxed. In studies on bulk materials or geometry-independent systems, calculations may not vary significantly; however, in nanometric systems – especially low-dimensional ones– the relaxation process becomes transcendent due to its impact on the final results. For the particular case of the monodimensional HAGNRs, a severely-relaxed HAGNR may not exist in nature and, for that reason, only partially-relaxed systems would make sense to be simulated; or may exist but in very particular cases after a certain fabrication process and in a determined surrounding (i.e., environment or setting).

## ACKNOWLEDGMENTS

We want to thank Miguel Pruneda Santos for fruitful discussions about TranSIESTA parameters. This work was supported by Spanish MICINN-FEDER project TEC2013-47283-R and by the European Union Seventh Framework Programme under grant agreement no 604391 Graphene Flagship. The calculations have been made in the UGRgridsupercomputation facilities (Alhambra supercomputation facilities now) in the University of Granada.